\documentclass[preprint,nofootinbib]{revtex4}

\usepackage{amsmath}
\usepackage{graphicx}

\newcommand{\beq}{\begin{equation}}
\newcommand{\eeq}{\end{equation}}
\newcommand{\beqa}{\begin{eqnarray}}
\newcommand{\eeqa}{\end{eqnarray}}
\newcommand{\no}{\nonumber}
\def\OMIT#1{{}}
\newcommand{\lsim}{\mathrel{\rlap{\lower4pt\hbox{\hskip1pt$\sim$}}
    \raise1pt\hbox{$<$}}}         
\newcommand{\gsim}{\mathrel{\rlap{\lower4pt\hbox{\hskip1pt$\sim$}}
    \raise1pt\hbox{$>$}}}         

\begin{document}

\vspace*{.0cm}

\title{Beyond MSSM Baryogenesis}

\author{Kfir Blum}\email{kfir.blum@weizmann.ac.il}
\affiliation{Department of Condensed Matter Physics,
  Weizmann Institute of Science, Rehovot 76100, Israel}
\author{Yosef Nir\footnote{The Amos de-Shalit chair of theoretical physics}}\email{yosef.nir@weizmann.ac.il}
\affiliation{Department of Particle Physics,
  Weizmann Institute of Science, Rehovot 76100, Israel\vspace*{3mm}}


\begin{abstract}
  Taking the MSSM as an effective low-energy theory, with a cut-off
  scale of a few TeV, can make significant modifications to the
  predictions concerning the Higgs and stop sectors. We investigate
  the consequences of such a scenario for electroweak baryogenesis. We
  find that the window for MSSM baryogenesis is extended and, most
  important, can be made significantly more natural. Specifically, it
  is possible to have one stop lighter than the top and the other
  significantly lighter than TeV simultaneously with the Higgs mass
  above the LEP bound. In addition, various aspects concerning CP
  violation are affected. Most notably, it is possible to have
  dynamical phases in the bubble walls at tree level, providing CP
  violating sources for Standard Model fermions.
\end{abstract}

\maketitle

\section{Introduction}
\label{sec:intro}
One of the strongest tests of the minimal supersymmetric standard
model (MSSM) comes from the fact that within this framework, the mass
of the lightest Higgs is bounded from above. Actually, the
experimental lower bound on the Higgs mass \cite{Yao:2006px} violates
the tree level bound and implies that, if the MSSM is indeed realized
in nature, stop-related loop corrections to the Higgs mass play an
important role. For these corrections to be significant, the stop
sector is required to exhibit special features: at least one of the
stop mass eigenstates should be rather heavy and/or left-right-stop
mixing should be substantial.

It is conceivable that additional new physics, beyond the MSSM, plays
a role in particle interactions not far above the TeV scale. Model
independent analyses of such scenarios were taken up in, for example,
Refs. \cite{Strumia:1999jm,Polonsky:2000aa,Brignole:2003cm,Casas:2003jx} and, more
recently, by Dine, Seiberg and Thomas [DST] in Ref.
\cite{Dine:2007xi}. DST extend the MSSM by adding non-renormalizable
terms that depend on the MSSM fields and are subject to the symmetries
of the MSSM. They find two such terms that could increase the Higgs
mass comparably to what loop corrections with heavy or mixed stops can
do. These works demonstrate the sensitivity of the mass spectrum of
the MSSM Higgs sector to changes in the quartic coupling.

Both the Higgs sector and the stop sector play a role in yet another
interesting aspect of the MSSM, and that is the cosmological scenario
of MSSM electroweak baryogenesis (EWBG) \cite{Cohen:1993nk}. In this
scenario, the two problems of the standard model baryogenesis -- the
fact that, with a single Higgs and taking into account experimental
constraints on its mass, the electroweak phase transition (EWPT)
cannot be strongly first order, and the fact that CP violation from
the Kobayashi-Maskawa phase is much too small -- are fixed by
supersymmetric particles and their interactions. In particular, a
light stop is crucial in making the EWPT strongly first order
\cite{Carena:1996wj,Delepine:1996vn}. It is therefore interesting to
ask how do the DST operators affect EWBG. In particular, since these
operators allow -- given a fixed Higgs mass -- lighter stops, do they
open up new regions in the MSSM parameter space for successful
baryogenesis?

In this work we focus on the EWPT and show that the parameter space
required for a strong EWPT does indeed change qualitatively by DST
operators. We discuss the implications of these qualitative features
on the naturalness of the required parameter space. The various issues
of calculating the baryon asymmetry of the universe (BAU) in a given
particle physics model are a subject of active research (see, for
example,
\cite{Konstandin:2004gy,Konstandin:2005cd,Cirigliano:2006wh,Fromme:2006wx}).
We do not carry out, at this stage, a detailed explicit calculation of
the BAU compatible with a low-cutoff MSSM, but rather point out
several modifications by which the mechanism responsible for the
produced baryon asymmetry may differ in the low-cutoff scenario from
the renormalizable MSSM case.

The narrowing EWBG-window in the MSSM parameter space has led many
authors to study extensions of the SM and of the MSSM
in this context. Some recent examples are
\cite{Kang:2004pp,Carena:2004ha,Huber:2006ma,Fromme:2006cm,Espinosa:2007qk}.
In most of these works, new particle degrees of freedom which are added
to the model modify the EWPT by actively participating in thermal
processes. Some previous studies have also considered the effect of
non-renormalizable corrections to the SM potential
\cite{Bodeker:2004ws,Grojean:2004xa,Delaunay:2007wb}. Our analysis
differs from previous studies in that it considers the most general
non-renormalizable correction to the MSSM scalar potential. We find
that there is no need for the beyond-MSSM (BMSSM) physics to actively
couple to the thermal plasma in order to get significant modifications
to various aspects of the EWBG.

The plan of this paper goes as follows. In Section \ref{sec:mssm} we
review the predictions of the MSSM for the lightest Higgs mass. In
Section \ref{sec:DST} we give the modifications to these predictions
due to the DST operators. In Section \ref{sec:V} we present the
modifications to the finite-temperature one-loop scalar potential
from the DST terms. Our main results are derived in Section
\ref{sec:impact} where we explain how the contributions of the
non-renormalizable terms enlarge the window for MSSM baryogenesis
and, in particular, relax the fine-tuning problem of this scenario.
The effects of these terms on the CP violating aspects of MSSM
baryogenesis are discussed in Section \ref{sec:cpv}: first, the
possibility of CP violating bubble wall profiles (Subsection
\ref{subsec:sponBG}) and, second, the modification to neutralino and
chargino currents (Subsection \ref{subsec:modSUSY}). We summarize
our results in Section \ref{sec:con}. In Appendix \ref{app:subs} we
clarify some subtleties related to substituting the dimensionful
parameters of the Higgs potential with measurable quantities in the
presence of the DST operators.

\section{The Lightest Higgs Mass in the MSSM}
\label{sec:mssm}
Within the MSSM, the lightest Higgs mass is bounded from
above. We write
\beq
m_h^2=m_h^{2({\rm tree})}+m_h^{2({\rm loop})}.
\eeq
The tree level contribution is given by
\beqa
m_h^{2({\rm tree})}&=&\frac12\left[m_Z^2+m_A^2
  -\sqrt{(m_A^2-m_Z^2)^2+4m_A^2m_Z^2\sin^22\beta}\right]\no\\
&\approx&m_Z^2-\frac{4m_Z^2m_A^2}{m_A^2-m_Z^2}\cot^2\beta,
\eeqa
where the second, approximate equality holds for large $\tan\beta$.
The most significant loop contribution is given by
\beqa
m_h^{2({\rm loop})}&=&\frac{3m_t^4}{4\pi^2
  v^2}\left[\ln\left(\frac{m_{\tilde t_1}m_{\tilde t_2}}{m_t^2}\right)
  +\frac{|X_t|^2}{m_{\tilde t_1}^2-m_{\tilde t_2}^2}
  \ln\left(\frac{m_{\tilde t_1}^2}{m_{\tilde t_2}^2}\right)\right.\no\\
&&\left.+\frac12\left(\frac{|X_t|^2}{m_{\tilde t_1}^2-m_{\tilde
        t_2}^2}\right)^2
  \left(2-\frac{m_{\tilde t_1}^2+m_{\tilde t_2}^2}{m_{\tilde
        t_1}^2-m_{\tilde t_2}^2}
    \ln\left(\frac{m_{\tilde t_1}^2}{m_{\tilde
          t_2}^2}\right)\right)\right],
\eeqa
where $X_t=A_t-\mu^*\cot\beta$.

The current lower bound from LEPII is $m_{h_{\rm SM}}\gsim114$ GeV,
well above the upper bound on the tree contribution, $m_h^{\rm
  (tree)}\leq m_Z$. This contribution by itself is maximized at
moderate to large $\tan\beta$. The top-stop loop correction should
be substantial. If stop mixing is small, $|X_t/m_{\tilde
  t_{1,2}}|^2\ll1$, the correction depends only on the logarithm of
the stop masses, so these must be rather heavy:
\beqa\label{eq:stsT}
(m_{\tilde t_1}m_{\tilde t_2})^{1/2}&\sim&
m_t\exp{\left[\frac{2\pi^2v^2\left(m_h^2-m_Z^2\right)}{3m_t^4}\right]}\no\\
&\sim&500\ {\rm GeV}\times \exp\left\{2.9\left[\left(\frac{m_h}{114\ {\rm
          GeV}}\right)^2-1\right]\right\}.
\eeqa
If, however, stop mixing is large, much lighter stops can still yield
large loop corrections.

\section{The DST operators and the light Higgs mass}
\label{sec:DST}
DST consider the following two non-renormalizable terms
\cite{Dine:2007xi}:
\beq\label{eq:wdst}
W_{\rm DST}=\frac{\lambda_1}{M}(H_uH_d)^2+\frac{\lambda_2}{M}{\cal
  Z}(H_uH_d)^2,
\eeq
where ${\cal Z}$ is s SUSY-breaking spurion:
\beq
{\cal Z}=\theta^2m_{\rm susy}.
\eeq
The first term in Eq. (\ref{eq:wdst}) is supersymmetric, while the
second breaks SUSY. In the scalar potential, the following quartic
terms are generated:
\beq\label{eq:dstsp}
\frac{2\mu^*\lambda_1}{M}H_uH_d(H_u^\dagger H_u+H_d^\dagger H_d)
-\frac{m_{\rm susy}\lambda_2}{M}(H_uH_d)^2.
\eeq
We define
\beq \epsilon_1\equiv\frac{\mu^*\lambda_1}{M},\ \ \
\ \ \ \epsilon_2\equiv-\frac{m_{\rm susy}\lambda_2}{M}.
\eeq
The two terms in Eq. (\ref{eq:dstsp}) contribute to the lightest Higgs
boson mass as follows:
\beqa\label{eq:mhdst}
m_h^{2{\rm (dst)}}&=&2v^2\left[\epsilon_{2r}-2\epsilon_{1r}\sin2\beta
  -\frac{2\epsilon_{1r}\sin2\beta(m_A^2+m_Z^2)+\epsilon_{2r}\cos^22\beta
    (m_A^2-m_Z^2)}{\sqrt{(m_A^2-m_Z^2)^2+4m_A^2m_Z^2\sin^22\beta}}\right]
\no\\
&\simeq&-16v^2\epsilon_{1r}\cot\beta\frac{m_A^2}{m_A^2-m_Z^2}+{\cal
    O}(\epsilon_i\cot^2\beta),
\eeqa
where $\epsilon_{ir}={\cal R}e(\epsilon_i)$. Eq. (\ref{eq:mhdst})
gives the leading $\epsilon_i$-related shift to $m_h$, as long as
$\cot\beta\not\ll\epsilon_i$. (Otherwise, ${\cal O}(\epsilon_i^2)$
corrections are equally or even more important.)

DST give the following numerical example. Take $m_{\tilde
  t_{1,2}}\simeq300\ GeV$ with small mixing, $X_t\simeq0$.  Then, at
moderate to large $\tan\beta$, we obtain $m_h\simeq100\ GeV$.  In the
same small mixing limit, and taking conservatively $m_A\gg m_Z$, the
additional correction (\ref{eq:mhdst}) can accommodate $m_h\gsim114\
GeV$ for $\epsilon_{1r}\cot\beta\lsim-0.006$ ({\it e.g.},
$\epsilon_{1r}=-0.06$ and $\tan\beta=10$). Assume now that the
right-handed stop mass is bounded from above:
$m_{\widetilde{t}_R}\lsim170$ GeV (the relevance of this bound to our
purposes will become clear below). With $\epsilon_1\cot\beta=-0.006$,
one obtains for the left-handed stop $m_{\widetilde{t}_L}\gsim530$
GeV. We learn that the heavy stop mass is pushed quite high. Setting,
however, $\epsilon_1=0$, would force it to much higher values:
$m_{\widetilde{t}_L}\gsim2$ TeV. Any further increase in the Higgs
mass would correspond to an exponential increase in the heavy stop
mass, implying fine tuning.

\section{The scalar potential}
\label{sec:V}
The MSSM finite-temperature effective potential was calculated by several groups
\cite{Espinosa:1993yi,Delepine:1996vn,Carena:1996wj,Espinosa:1996qw,Bodeker:1996pc,Losada:1996ju,Farrar:1996cp,de
Carlos:1997ru,Cline:1998hy,Laine:1998vn,Davidson:1999ii}. Two-loop results are significant in improving the
one-loop calculations and, as demonstrated by non-perturbative analyses, provide a rather accurate description
of the full result \cite{Laine:2000rm,Csikor:2000sq}. The qualitative changes that follow from adding the DST
operators to the MSSM potential can, however, be well understood without the two-loop improvement. In fact, the
effect which we find most significant is a zero-temperature effect which leaves the thermal computation all but
idle. In this work we therefore employ the one-loop analysis. We leave a detailed quantitative discussion of
the modified MSSM parameter space to more sophisticated, two-loop computations.

We represent the Higgs fields in the following way:
\beqa\label{H} H_d=\left(\begin{array}{c} H^0_d \\
    H_d^- \end{array}\right)
=\left(\begin{array}{c} \phi_1+\frac{H^0_{dr}+iH^0_{di}}{\sqrt{2}}
    \\
H_d^- \end{array}\right), \ \ \ \  H_u=\left(\begin{array}{c} H_u^+
\\ H^0_u \end{array}\right)=\left(\begin{array}{c} H_u^+ \\
\phi_2+\frac{H^0_{ur}+iH^0_{ui}}{\sqrt{2}} \end{array}\right)
\eeqa
with the VEVs $\phi_1=\langle H^0_d\rangle$ and $\phi_2=\langle
H^0_u\rangle$.
The DST terms, Eq. (\ref{eq:dstsp}), contribute the following
dimension-four (but {\it effective dimension}-five \cite{Dine:2007xi})
terms to the tree level effective potential:
\beq\label{eq:V_DST}
V_{\rm dst}=-2\left(|\phi_1|^2+|\phi_2|^2\right)\left[
  \epsilon_{1}\phi_1\phi_2+{\rm h.c.}\right] +
\left[\epsilon_2\left(\phi_1\phi_2\right)^2+{\rm h.c.}\right].
\eeq

The dimension-six terms,
\beq
V_{\rm dst}^{(6)}=
\left|{2\epsilon_1}/{\mu}\right|^2|\phi_1\phi_2|^2
\left(|\phi_1|^2+|\phi_2|^2\right),\no
\eeq
are important if the quartic couplings by themselves
destabilize the potential. This has been demonstrated for a SM-like
Higgs \cite{Bodeker:2004ws,Grojean:2004xa,Delaunay:2007wb}. We focus,
however, on the case where the DST operators actually raise the
quartic coupling above the level predicted by the MSSM. In this case,
the dimension-six terms can be neglected if they are suppressed in
comparison to the dimension-four terms:
\beq
\epsilon_1\left|{\phi}/{\mu}\right|^2\cot\beta\ll1.\no
\eeq
The dimension-six operators also enter the potential via
field-dependent thermal corrections. In our case the leading
high-temperature terms due to such corrections, which are absent
otherwise, are of the form
$\left|\frac{\epsilon_1}{\mu}\cot\beta\right|^2
\left(\phi^4T^2+\phi^2T^4\right)$ (we omit here numerical factors).
For $|\mu|\gsim100$ GeV, and within the temperature range of interest,
these contributions are suppressed in comparison with the usual
$\phi^2T^2$ thermal terms which are accompanied by much larger
coefficients.

Including the DST terms of Eq. (\ref{eq:V_DST}), the one-loop
effective potential for the Higgs scalars at finite temperature is
given by
\beqa\label{eq:veff}
V&=&m^2_1|\phi_1|^2+m^2_2|\phi_2|^2
-\left(m^2_{12}\phi_1\phi_2+{\rm h.c.}\right)+
\frac{g^2+g'^2}{8}\left(|\phi_1|^2-|\phi_2|^2\right)^2\no\\
&-&2\left(|\phi_1|^2+|\phi_2|^2\right)
\left[\epsilon_{1}\phi_1\phi_2+{\rm h.c.}\right]+
\left[\epsilon_2\left(\phi_1\phi_2\right)^2+{\rm h.c.}\right]\no\\
&+&\sum_{i=\{{\rm dof}\}}\frac{n_i m_i^4(\phi)}{64\pi^2}\left[
    \ln\left(\frac{m_i^2(\phi)}{Q^2}\right)-\frac32\right]\no\\
  &+&\sum_{i=\{{\rm dof}\}}n_i\frac{T^4}{2\pi^2}J_i\left(
    \frac{m_i^2(\phi)}{T^2}\right)+
  \sum_{i=\{{\rm sca}\}}\frac{n_i T}{12\pi}\left[
   m_i^3(\phi)-\bar m_i^3(\phi,T)\right].
\eeqa
The first line corresponds to the tree-level MSSM terms. The second
line corresponds to the tree-level DST contributions. The third line
is the zero-temperature one-loop contribution. The summation goes over
\beq
\{{\rm dof}\}=\{t,b,\tilde t_{1,2},\tilde b_{1,2},H_e,H_o,H_c,
W_T,Z_T,\gamma_T,W_L,Z_L,\gamma_L\},
\eeq
with
\beqa
n_t&=&n_b=-12,\ n_{\tilde t_{1,2}}=n_{\tilde b_{1,2}}=6,\no\\
n_{H_e}&=&n_{H_o}=2,\ n_{H_c}=4,\no\\
n_{W_T}&=&4,\ n_{Z_T}=n_{\gamma_T}=2,\ n_{W_L}=2,\
n_{Z_L}=n_{\gamma_L}=1.
\eeqa
Here $H_e$ and $H_o$ refer to, respectively, the two CP-even and two
CP-odd neutral Higgs bosons; $H_c$ are the charged Higgs bosons;
sub-indices $T$ and $L$ stand for, respectively, transverse and
longitudinal. The fourth line is the finite-temperature
contribution. The $J_i$ functions are defined by
\beq
J_i(r)=\int_0^\infty dx\ x^2\ln[1-(-1)^{2s_i}e^{-\sqrt{x^2+r}}].
\eeq
The last term corresponds to daisy improvement. The masses $\bar
m_i^2(\phi,T)$ are the field- and temperature-dependent eigenvalues of
the mass matrices with first-order thermal masses included. We use the
conventions of Ref. \cite{de Carlos:1997ru}, where the reader is
referred to for further details. The summation is over
\beq
\{{\rm sca}\}=\{\tilde
t_{1,2},\tilde b_{1,2},H_e,H_o,H_c,W_L,Z_L,\gamma_L\}.
\eeq

\section{The electroweak phase transition}
\label{sec:impact}
The most important effect that we find comes from the tree level
change in the quartic couplings of the scalar potential of Eq.
(\ref{eq:veff}). The most significant consequence of this effect is
that it allows a strongly first-order EWPT with a relatively light
$\widetilde{t}_L$. In this section, we analyze this effect.

The region in MSSM parameter space which is compatible with a strong
enough first-order phase transition (the \emph{MSSM window}) has two
distinctive characteristics (see
\cite{Cline:2006ts,Quiros:2007zz,Buchmuller:2007fd} and references
therein):
\begin{enumerate}
\item {A light, (mostly) right-handed stop:
\beq\label{eq:mtcond}
m_{\widetilde{t}_R}\lsim m_t;
\eeq}
\item {A light Higgs, close to the LEP lower bound:
\beq\label{eq:mhcond}
m_h\approx115 \ GeV.
\eeq}
\end{enumerate}
In order to understand how this window is affected by the DST
operators, we now explain how these constraints come about.

The condition for the sphaleron processes in the broken phase not to
erase the baryon asymmetry that is produced along the expanding
bubble wall reads \cite{Shaposhnikov:1986jp}
\beq\label{eq:vcTc}
\frac{\sqrt{2}v_c}{T_c}\gsim1.
\eeq
Here $v_c=v(T_c)$ and $T_c$ are the Higgs VEV and the temperature at
the instance in which the symmetric and the asymmetric vacua become
degenerate. The normalization is such that $v_0=v(T=0)=174 \ GeV$.

The light stop constraint, Eq. (\ref{eq:mtcond}), comes from the need to reduce thermal screening for at least
one scalar which has a large coupling to the Higgs field \cite{Carena:1996wj,Delepine:1996vn}. EW precision
measurements can be accommodated more easily if this light stop is dominantly `right-handed'. Let us focus on
the case of large but finite $m_A^2\gg m_Z^2$, relevant to the analysis of Ref. \cite{Dine:2007xi}. The
minimization of the potential reduces in this case to a one dimensional problem
\cite{Carena:1996wj,Quiros:1999jp}, yielding
\beq\label{eq:cond1loop}
\frac{v_c}{T_c}\approx\frac{E}{\lambda}.
\eeq
Here $E$ is the coefficient of the cubic (barrier) term, and $\lambda$
is the effective quartic coupling for the light Higgs. If the soft
mass-squared of $\widetilde{t}_R$ is chosen negative such that it
cancels exactly the thermal mass at the critical temperature, one has
\beq\label{eq:E} E\approx\frac{h_t^3\sin^3\beta\left(1-X_t^2/m_Q^2\right)^{3/2}}{2\pi}.
\eeq
For small stop mixing, $X_t^2/m_Q^2\ll1$, $E$ can be of order $0.1$
and thus an order of magnitude larger than the SM contribution due to
transverse gauge bosons, $E_{\rm SM}\sim0.01$. Eq. (\ref{eq:E}) biases
the MSSM window towards small stop mixing regions. Most importantly,
the requirement of negative $m_U^2$ forces $m_{\widetilde{t}_R}<m_t$.
Within one-loop analysis, one must in fact impose a rather strong
constraint, $m_U^2\sim-\left(80\ GeV\right)^2$ or equivalently
$m_{\widetilde{t}_R}\sim150$ GeV, to obtain a strong enough PT.

Two-loop calculations (see, for example,
\cite{Davidson:1999ii,Quiros:1999jp,Cline:2006ts} and references
therein) extend the window by correcting Eq. (\ref{eq:cond1loop}):
\beq\label{eq:cond2loop} \frac{v_c}{T_c}\approx\frac{E}{2\lambda}
+\sqrt{\frac{E^2}{4\lambda^2}+\frac{c_2}{\lambda}}, \eeq where $c_2$
is the coefficient of the generic two-loop correction,
\[\Delta V^{\rm (2-loop)}\approx-c_2T^2\phi^2\ln\frac{\phi}{T}.\]
Eq. (\ref{eq:cond2loop}) explains how two-loop corrections make room for some stop mixing and relax the upper
bound on $m_U^2$. However, sizeable positive values of $m_U^2$ or large mixing are still forbidden, as they
directly decrease $E$. In addition, the effective quartic coupling $\lambda$ remains in the denominator of
(\ref{eq:cond2loop}).

The light Higgs constraint, Eq. (\ref{eq:mhcond}), can also be
understood from the previous discussion. Since $\lambda$ is
proportional to the zero temperature value of the Higgs mass-squared
(up to a weak logarithmic dependence which distinguishes $\lambda$ of
Eqs. (\ref{eq:cond1loop},\ref{eq:cond2loop}) from the zero-temperature
$\lambda$), $\lambda\approx{m_h^2}/{(2v_0^2)}$, the constraint of a
light Higgs turns out to be a zero-temperature effect in the MSSM.

In fact, Refs. \cite{Cline:1998hy,Cline:2006ts} note that the
requirement of a light Higgs in the MSSM window does not arise
directly from the two-loop thermal calculation. Instead, it is a
consequence of the theoretical ``upper bound''
$m_{\widetilde{t}_L}\lsim$ TeV, coming from fine-tuning
considerations. The question is how can an increase in the Higgs mass
above the LEP bound be accounted for in the MSSM, where the stop-Higgs
relation of Eq. (\ref{eq:stsT}) holds. Within the MSSM window,
that is when  Eq. (\ref{eq:mtcond}) is obeyed, a corresponding
(exponentially large!) increase of $m_{\widetilde{t}_L}$ is
required. In contrast, outside the baryogenesis window, the task can
be shared among the two stops. Two-loop analyses
\cite{Cline:1998hy,Cline:2006ts} show that this naive argument is
qualitatively correct, though quantitatively the bound is somewhat
weaker than what follows from Eq. (\ref{eq:stsT}). Within the MSSM
window, they obtain
\beq\label{eq:mQCline}
m_Q\cong 100 \
GeV\times\exp{\left[0.11\left({m_h}{[GeV]}-85.9\right)\right]}.
\eeq
This strong constraint implies, for example, that in order to account
for $m_h\sim120$ GeV, one must take $m_{\widetilde{t}_L}\sim4
\ TeV$ within the MSSM window, while outside this window one can
have $m_{\widetilde{t}_L}\sim m_{\widetilde{t}_R}\sim680 \ GeV$, a
much less fine-tuned situation.

\begin{figure}[h]\center
\includegraphics[width=120mm]{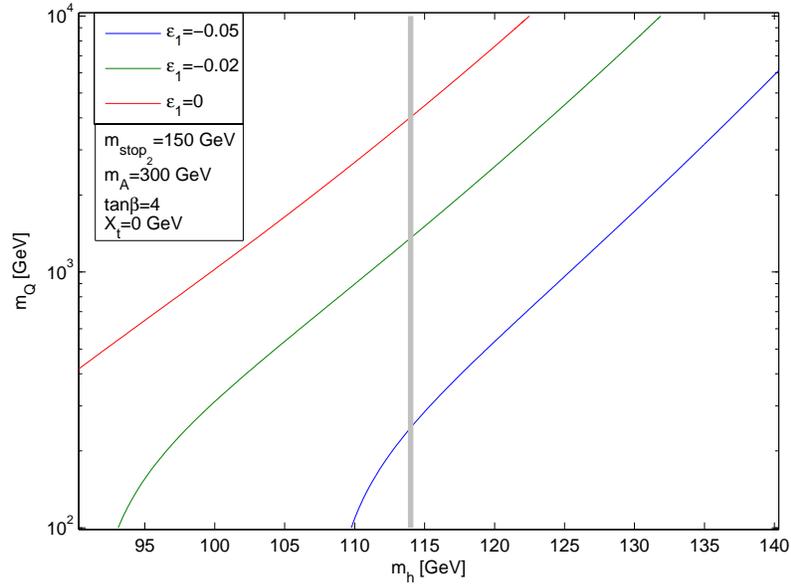}
\caption{The soft SUSY-breaking mass of $\widetilde{t}_L$, $m_Q$,
  versus the light Higgs mass, $m_h$, for various values of
  $\epsilon_1$. Other relevant parameters are fixed at $X_t=0$,
  $m_{\widetilde{t}_R}=150$ GeV, $m_A=300$ GeV and $\tan\beta=4$. The
  gray line marks the LEP lower bound, $m_h=114$ GeV. }
\label{fig:HiggsMassStopEps}
\end{figure}

This is the point where the effect of the DST corrections
(\ref{eq:mhdst}) is most significant. The DST operators modify the
quartic coupling at zero-temperature, and consequently the task of
stabilizing the potential can be shared between them and the stop
sector. This adds a new twist to a well-recognized fact: The smallness
of the quartic coupling in the tree level MSSM makes the spectrum
extremely sensitive not only to quantum corrections but also to
non-renormalizable contributions. However, in contrast to the loop
corrections which require in this case a strong fine tuning, the
low-cutoff corrections relieve this tuning by allowing a light soft
mass for the stop \cite{Casas:2004uu}. We demonstrate this effect in
Fig.  \ref{fig:HiggsMassStopEps}, where we plot curves of constant
$\epsilon_1$ in the $m_h-m_Q$ plane. For the purpose of illustration,
we use vanishing stop mixing, $X_t=0$, and a light right-handed stop,
$m_{\widetilde{t}_R}=150 \ GeV$, consistent with a strongly
first-order EWPT at one-loop. One learns from this figure that even a
modest non-renormalizable correction, $\epsilon_{1r}\sim-0.05$,
suffices to reduce fine tuning in the Higgs sector by two orders of
magnitude, making a sizable part of the traditional MSSM window
significantly more natural.

\begin{figure}[h]\center
\includegraphics[width=120mm]{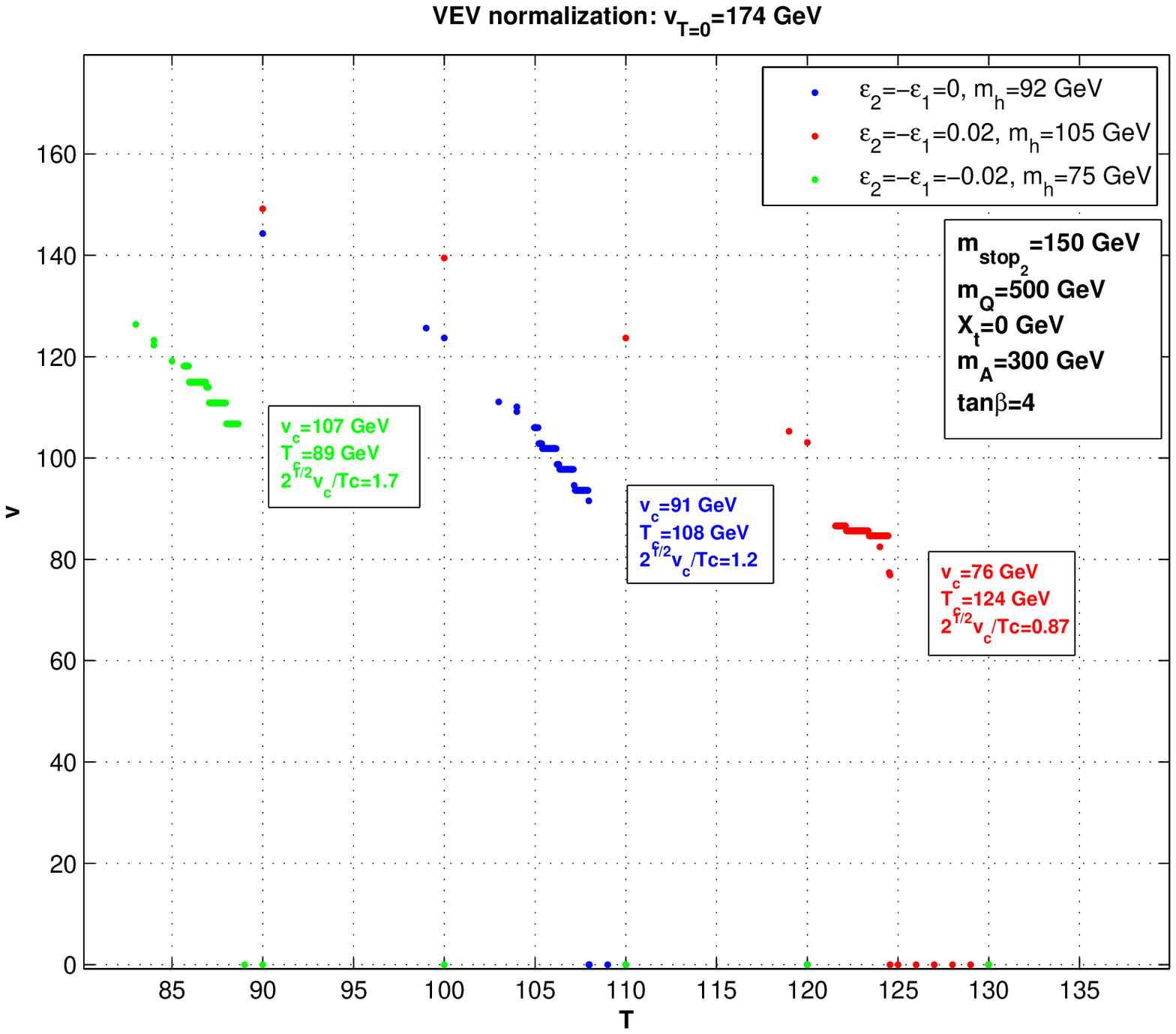}
\caption{The Higgs VEV at the lowest minimum as a function of
  temperature for various values of $m_h$, $\epsilon_1$ and
  $\epsilon_2$. Other relevant parameters are fixed at $X_t=0$,
  $m_{\widetilde{t}_R}=150$ GeV, $m_Q=500$ GeV, $m_A=300$ GeV and
  $\tan\beta=4$. The different sets of ($m_h,\epsilon_1,\epsilon_2$)
  all yield the same $m_h^2(v_c/T_c)$ to within about $5\%$.}
\label{fig:PT_profile_FixLightStop}
\end{figure}

In Fig. \ref{fig:PT_profile_FixLightStop} we evaluate the one loop
potential, in order to support our statement that the opening of the
MSSM window
for a lighter left-handed stop is a zero-temperature effect. We
present in the figure the order parameter
$v(T)=\sqrt{|\phi_1|^2+|\phi_2|^2}$ at the true vacuum as a function
of temperature for different values of $\epsilon$. (Notice the
first-order nature of the EWPT, even at one loop, for the selected set
of parameters.) Keeping the left-handed stop heavy and decoupled, we
see that the effect of varying $\epsilon$ is mainly to shift the
resulting values of the critical VEV and temperature such that
$({v_c}/{T_c})\cdot m_h^2\approx$ const. This confirms the simple
expectation of Eq. (\ref{eq:cond1loop}), and shows that a change in
${v_c}/{T_c}$ due to the DST operators is simply a result of the
zero-temperature change in Higgs mass. Two-loop and lattice
calculations then imply that a strongly first-order EWPT can occur
even if $m_h$ is well above the LEP bound.

Let us conclude this section by summarizing the qualitative picture
that arises from our analysis. The main results can be understood
from Eqs. (\ref{eq:vcTc}) and (\ref{eq:cond1loop}). The resulting
constraints on the relevant supresymmetric parameters can be
schematically presented as follows:
\beq
\frac{E(m_{\tilde t_R})}{\lambda(m_{\tilde t_R},m_{\tilde
    t_L},\epsilon_i)}\geq 1.
\eeq
Given that there is an {\it experimental} lower bound on $\lambda$,
the requirement of strongly first order EWPT translates into a lower
bound on $E$ which, in our framework, requires $m_{\tilde t_R}$ to be
within the narrow range between the lower bound (coming from direct
searches and/or the requirement that there is no color breaking) and
(roughly) $m_t$. Since $E$ does not dpened directly on $\epsilon_i$,
this constraint is hardly affected by extending the MSSM with
non-renormalizable terms. Thus, the main effect of $\epsilon_i$ is
that it can be combined with $m_{\tilde t_L}$ to render $\lambda$
close to the lower bound (with $m_{\tilde t_R}$ almost fixed). In
particular, negative values of $\epsilon_1$ allow lower values of
$m_{\tilde t_L}$ compared to the MSSM. This is the content of Fig.
\ref{fig:HiggsMassStopEps}. We expect other, smaller effects, on the
allowed range of $\tan\beta$ and $X_t$, but to quantify them we need
the full two-loop calculation.

\section{CP violation}
\label{sec:cpv}
The non-renormalizable terms affect not only the phase transition,
but also the CP violation that is relevant to baryogenesis. First,
they induce CP violating bubble wall profiles and by that allow the
third generation fermions to directly produce some baryon asymmetry.
Second, they modify the chargino and neutralino currents. These two
effects are described in the two respective subsections.

\subsection{CP violation in bubble wall}
\label{subsec:sponBG}
In the tree level potential of the MSSM, the $m_1^2$ and $m_2^2$
parameters are real, while the $m_{12}^2$ parameter can have an
arbitrary phase. However, one may use global field redefinitions to
make $m_{12}^2$ real and positive. In this basis, $\phi_1$ and
$\phi_2$ are real and positive throughout the phase transition. Thus
one arrives at the well known conclusion that, within the MSSM, CP
violation in bubble walls is insignificant
\cite{Cohen:1991iu,Abel:1992za,Huber:1999sa} (see, however,
\cite{Comelli:1993ne}).

Adding the DST operators changes this picture. Let us define
\beqa\label{eq:vdef}
v^2&=&|\phi_1|^2+|\phi_2|^2, \no\\
\tan\beta&=&|\phi_1/\phi_2|,\ \ s=\sin\beta,\ c=\cos\beta,\no\\
e^{-i\theta}&=&v^2sc/(\phi_1\phi_2).
\eeqa
Note that $v$ and $\beta$ (and $\theta$) parameterize the VEVs at the
{\it finite}-temperature vacuum. The phase dependent part in the tree
level potential is
\beq\label{eq:spontBG}
-\left[\left(m^2_{12}+2v^2\epsilon_{1}\right)e^{-i\theta}-
  \epsilon_2v^2sce^{-2i\theta}+{\rm h.c.}\right]v^2sc.
\eeq
In contrast to the MSSM scenario, it is not possible -- even at tree
level -- to globally define the phases of $\phi_i$ such that the
coefficient in square brackets in Eq. (\ref{eq:spontBG}) is maximized
for varying $v$.  During the PT, the phase of $\phi_1\phi_2$ aligns
dynamically, inducing an energy gap between left and right chiral
components of fermions \cite{Cohen:1991iu}.  This situation resembles
the case of the two-Higgs doublet model \cite{Fromme:2006cm}. It is
different from the MSSM scenario in that, for example, the BAU can be
generated through top or tau rather than chargino or neutralino
currents.

The produced BAU due to a varying complex phase in the bubble wall is
proportional to an integral over the gradient of the phase across the
wall. To find the dynamical phase profile requires obtaining the
tunneling solution, as in \cite{Huber:1999sa}. Here we use Eq.
(\ref{eq:spontBG}) to estimate the total variation of
$\theta$ between the symmetric and broken vacua, based on potential
energy alone. Let us neglect for now the effect of $\epsilon_2$. In
this limit, $\theta$ is given by
\beq
\theta=\arg{\left(m_{12}^2+2v^2\epsilon_1\right)}.\no
\eeq
Adopting a basis for the symmetric vacuum in which $m_{12}^2$ is real
and positive, we get, to leading order in $\epsilon_1$,
\beq\label{eq:Dphi}
\Delta\theta\approx2\epsilon_{1i}\left(\frac{v_c}{m_A}\right)^2\tan\beta
\eeq
We substitute here $m_{12}^2\approx\frac{1}{2}m_A^2\sin2\beta$, which
holds at tree level and to zeroth order in $\epsilon$ (see Appendix
\ref{app:subs}). Note that our approximations here and in Section
\ref{sec:DST} hold for $(m_A/v_0)^2>2\epsilon\tan\beta$. Thus the phase
cannot be large, $\Delta\theta<0.3$.

The tree level CP violation at the bubble walls triggers some amount
of baryogenesis, through varying complex phases in the field-dependent
masses of SM particles, like top and bottom quarks or tau leptons.
Which of these plays the most significant role depends on the
specifics of the scenario under consideration, particularly the value
of $\tan\beta$.

\subsection{SUSY particle currents}
\label{subsec:modSUSY}
In the renormalizable MSSM, mass eigenmodes in the chargino and
neutralino sectors develop time-varying complex phases due to the
variation of EW breaking, real, off-diagonal terms in the associated
mass matrices. CP violation is provided by complex-valued SUSY and
soft SUSY-breaking parameters. DST operators affect this computation,
even in case that they do not introduce any new phases.

We continue to use the parametrization of Eq. (\ref{eq:vdef}). The
neutralino $\widetilde{N}$ and the chargino $\widetilde{C}$ mass
matrices get contributions from the dimension-five DST operator
\cite{Dine:2007xi}:
\beq\label{eq:dst5}
\frac{\epsilon_1}{\mu^*}\left[2(H_uH_d)(\widetilde{H_u}\widetilde{H_d})
+2(\widetilde{H_u}H_d)(H_u\widetilde{H_d})+(H_u\widetilde{H_d})(H_u\widetilde{H_d})
+(\widetilde{H_u}H_d)(\widetilde{H_u}H_d)\right]+{\rm h.c.},
\eeq
arising from the superpotential of Eq. (\ref{eq:wdst}). In the
gauge eigenstate basis, one obtains \cite{Martin:1997ns}:
\beq\label{eq:MNino}
M_{\widetilde{N}}=\left(\begin{array}{cccc} M_1 & 0 &
    -\frac{g'\phi_1}{\sqrt{2}} & \frac{g'\phi_2}{\sqrt{2}}\\
    0 & M_2 & \frac{g\phi_1}{\sqrt{2}} & -\frac{g\phi_2}{\sqrt{2}}\\
    -\frac{g'\phi_1}{\sqrt{2}} & \frac{g\phi_1}{\sqrt{2}} & 0 & -\mu\\
    \frac{g'\phi_2}{\sqrt{2}} & -\frac{g\phi_2}{\sqrt{2}} & -\mu & 0
\end{array}\right) -
\frac{\epsilon_1}{\mu^*}\left(\begin{array}{cccc} 0  & 0 & 0 & 0 \\
  0 & 0 & 0 & 0 \\ 0 & 0 & \phi_2^2 & 4\phi_1\phi_2\\
0 & 0 & 4\phi_1\phi_2 & \phi_1^2 \end{array}\right),
\eeq
\beq\label{eq:MCino}
M_{\widetilde{C}}=\left(\begin{array}{cc} M_2 & g\phi_2 \\
    g\phi_1 & \mu \end{array}\right)
+ \frac{2\epsilon_1}{\mu^*}\phi_1\phi_2
\left(\begin{array}{cc} 0 & 0 \\ 0 & 1 \end{array}\right).
\eeq
Apart from modifications of the spectrum, the DST corrections
introduce position dependence into the $\widetilde{H}-\widetilde{H}$
entries of the matrices. Position dependence of the diagonal
elements is by itself a new effect. The gradients of DST-induced
entries are, however, suppressed by factors of ${\cal
O}(\epsilon\phi/\mu)$, small compared to the usual suppression of
${\cal O}(g)$.

\section{Discussion}
\label{sec:con}
The tree-level renormalizable scalar potential of the MSSM yields an
upper bound on the mass of the lightest Higgs boson which is
experimentally known to be violated. Loop-corrections involving the
top quarks and squarks can (still?) relax the theoretical bound
sufficiently, at the cost of some fine-tuning. Non-renormalizable
terms, suppressed by a cut-off scale in the few TeV range, can
similarly relax the bound without, however, fine-tuning
\cite{Dine:2007xi}. We analyzed the consequences of such beyond-MSSM
(BMSSM) effects on electroweak baryogenesis (EWBG). We find that, even
in this regime, the non-renormalizable terms may easily alter some of
the principal features of the so called ``window'' for EWBG in the
MSSM.

In particular, the fine-tuning that arises if the required increase of
the quartic Higgs coupling is attributed solely to stop-related loop
corrections, becomes quite strong in the MSSM window for
baryogenesis. The reason is that, to have a large enough cubic term in
the scalar potential as necessary for strongly first-order phase
transition, the mass of the lighter stop must be below the top mass
which, in turn, requires that the heavy stop mass is in the few
TeV range. In our framework, however, the task of increasing the
quartic coupling can be shared between the stop-related
loop-corrections and the non-renormalizable contributions. This allows
a strongly first-order phase transition simultaneously with a Higgs
mass above the LEP bound with the heavier stop mass well below
TeV. Thus, the MSSM window is extended and, most significantly, made
natural.

Additional relevant consequences of the non-renormalizable terms
concern CP violation. The new operators provide new sources of CP
violation. These make a qualitative change in the picture of CP
violation. Unlike in the MSSM, it is impossible in general to choose
a phase convention whereby the relative phase between the two Higgs
VEVs vanishes all along the bubble wall. The interactions of the
third generation fermions -- top, bottom and tau -- with the bubble
wall could thus contribute significantly to the generation of the
baryon asymmetry. Finally, the chargino and neutralino currents,
which within the MSSM are usually responsible for the baryogenesis,
are modified in a qualitatively interesting way, though the
quantitative effects are parametrically suppressed and may turn out
small.

We conclude that BMSSM effects, which may become
necessary to give a consistent picture of the Higgs and stop sectors,
can play a significant role also in supersymmetric baryogenesis. In
particular, the BMSSM window for baryogenesis allows for parameters
that are significantly more natural than those of the MSSM
baryogenesis.

\section*{Acknowledgments}
We are grateful to Michael Dine and Ann Nelson for useful discussions
and to Michael Peskin, Scott Thomas and, in particular, Jose Espinosa,
for helpful correspondence. This research is supported
by the Israel Science Foundation founded by the Israel Academy of
Sciences and Humanities, the United States-Israel Binational Science
Foundation (BSF), Jerusalem, Israel, the German-Israeli foundation for
scientific research and development (GIF), and the Minerva Foundation.

%
%
\appendix

\section{Substituting $m_1^2,m_2^2,m_{12}^2$}
\label{app:subs}
The effective potential of Eq. (\ref{eq:veff}) depends on three
dimensionful quantitities: $m_1^2,m_2^2$ and $m_{12}^2$.  These
parameters need to be replaced by measurable quantities, for which we
choose $m_Z,m_A$ and $\tan\beta$ at the zero-temperature vacuum:
\beqa\label{eq:A1}
m_{12}^2&=&\frac12 m_A^2 s_{2\beta}-2\epsilon_{1r}
v^2+2\epsilon_{2r} v^2 s_{2\beta} +\frac{3g^2 m_t^2 A_t\mu}{32\pi^2
  s^2_\beta m_W^2}g(m_{\tilde t_1}^2,m_{\tilde t_2}^2) +\frac{3g^2
  m_b^2 A_b\mu}{32\pi^2 c^2_\beta
  m_W^2}g(m_{\tilde b_1}^2,m_{\tilde b_2}^2),\no\\
g(m_1^2,m_2^2)&=&\frac{m_1^2[\ln(m_1^2/Q^2)-1]
  -m_2^2[\ln(m_2^2/Q^2)-1]}{m_1^2-m_2^2},\no\\
m_1^2&=&m_{12}^2\tan\beta-\frac12
m_Z^2 c_{2\beta}+2\epsilon_{1r}
v^2(\tan\beta+s_{2\beta})-2\epsilon_{2r} v^2 s^2_\beta\no\\
&&-\frac{1}{64\pi^2}\left\{\sum_{i=\{{\rm
      dof}\}}n_i\frac{m_i^2(\phi)}{\phi_1}\frac{\partial
    m_i^2(\phi)}{\partial\phi_1}\left[
    \ln\left(\frac{m_i^2(\phi)}{Q^2}\right)-1\right]\right\}_{v,\beta},\no\\
m_2^2&=&m_{12}^2\cot\beta+\frac12 m_Z^2 c_{2\beta}+2\epsilon_{1r}
v^2(\cot\beta+s_{2\beta})-2\epsilon_{2r} v^2 c^2_\beta\no\\
&&-\frac{1}{64\pi^2}\left\{\sum_{i=\{{\rm
      dof}\}}n_i\frac{m_i^2(\phi)}{\phi_2}\frac{\partial
    m_i^2(\phi)}{\partial\phi_2}\left[
    \ln\left(\frac{m_i^2(\phi)}{Q^2}\right)-1\right]\right\}_{v,\beta}.
\eeqa
The values of $v,\beta$ in Eqs. (\ref{eq:A1}) refer to the
zero-temperature vacuum.

The basis in which we define the phases of $\epsilon$ above is defined
by having $m_{12}^2$ real and positive. Adopting the conventions of
Eq. (\ref{eq:vdef}), the $\theta$-dependent part in the tree-level
potential is approximately given by
\beq\label{eq:theta}-2v^4\cot\beta \cdot
{\cal R}e\left[\left(\frac{m_{12}^2}{v^2}
    +2\epsilon_{1}-\cot\beta\epsilon_2e^{-i\theta}\right)
  e^{-i\theta}\right].
\eeq
For $m_{12}^2/v^2\gg2\epsilon$, there is a small $\theta$
solution which minimizes (\ref{eq:theta}):
\beq \theta\approx\frac{v^2}{m_{12}^2}
\left(2\epsilon_{1i}-\cot\beta\epsilon_{2i}\right),\no \eeq
to leading order in $\epsilon$.  For this solution, we can treat
$\phi_{1,2}$ as positive numbers and replace the real part in
(\ref{eq:theta}) by absolute value:
\beq
-2m^2_{12}\phi_1\phi_2-4\epsilon_{1r}
\left(\phi_1^2+\phi_2^2\right)\phi_1\phi_2+
2\epsilon_{2r}\phi_1^2\phi_2^2,\no
\eeq
which is used in Eqs. (\ref{eq:A1}).




\begin{thebibliography}{99}

\bibitem{Yao:2006px}
  W.~M.~Yao {\it et al.}  [Particle Data Group],
  ``Review of particle physics,''
  J.\ Phys.\ G {\bf 33}, 1 (2006) and 2007 partial update for the 2008
  edition.

\bibitem{Strumia:1999jm}
  A.~Strumia,
  ``Bounds on Kaluza-Klein excitations of the SM vector bosons from
  electroweak tests,''
  Phys.\ Lett.\  B {\bf 466}, 107 (1999)
  [arXiv:hep-ph/9906266].

\bibitem{Polonsky:2000aa}
  N.~Polonsky,
  ``The scale of supersymmetry breaking as a free parameter,''
  Nucl.\ Phys.\ Proc.\ Suppl.\  {\bf 101}, 357 (2001)
  [arXiv:hep-ph/0102196].

\bibitem{Brignole:2003cm}
  A.~Brignole, J.~A.~Casas, J.~R.~Espinosa and I.~Navarro,
  ``Low-scale supersymmetry breaking: Effective description, electroweak
  breaking and phenomenology,''
  Nucl.\ Phys.\  B {\bf 666}, 105 (2003)
  [arXiv:hep-ph/0301121].

\bibitem{Casas:2003jx}
  J.~A.~Casas, J.~R.~Espinosa and I.~Hidalgo,
  ``The MSSM fine tuning problem: A way out,''
  JHEP {\bf 0401}, 008 (2004)
  [arXiv:hep-ph/0310137].

\bibitem{Dine:2007xi}
  M.~Dine, N.~Seiberg and S.~Thomas,
  ``Higgs Physics as a Window Beyond the MSSM (BMSSM),''
  Phys.\ Rev.\  D {\bf 76}, 095004 (2007)
  [arXiv:0707.0005 [hep-ph]].

\bibitem{Cohen:1993nk}
  A.~G.~Cohen, D.~B.~Kaplan and A.~E.~Nelson,
  ``Progress in electroweak baryogenesis,''
  Ann.\ Rev.\ Nucl.\ Part.\ Sci.\  {\bf 43}, 27 (1993)
  [arXiv:hep-ph/9302210].

\bibitem{Carena:1996wj}
  M.~S.~Carena, M.~Quiros and C.~E.~M.~Wagner,
  ``Opening the Window for Electroweak Baryogenesis,''
  Phys.\ Lett.\  B {\bf 380}, 81 (1996)
  [arXiv:hep-ph/9603420].

\bibitem{Delepine:1996vn}
  D.~Delepine, J.~M.~Gerard, R.~Gonzalez Felipe and J.~Weyers,
  ``A light stop and electroweak baryogenesis,''
  Phys.\ Lett.\  B {\bf 386}, 183 (1996)
  [arXiv:hep-ph/9604440].



\bibitem{Konstandin:2004gy}
  T.~Konstandin, T.~Prokopec and M.~G.~Schmidt,
  ``Kinetic description of fermion flavor mixing and CP-violating sources  for
  baryogenesis,''
  Nucl.\ Phys.\  B {\bf 716}, 373 (2005)
  [arXiv:hep-ph/0410135].

\bibitem{Konstandin:2005cd}
  T.~Konstandin, T.~Prokopec, M.~G.~Schmidt and M.~Seco,
  ``MSSM electroweak baryogenesis and flavour mixing in transport  equations,''
  Nucl.\ Phys.\  B {\bf 738}, 1 (2006)
  [arXiv:hep-ph/0505103].

\bibitem{Cirigliano:2006wh}
  V.~Cirigliano, M.~J.~Ramsey-Musolf, S.~Tulin and C.~Lee,
  ``Yukawa and tri-scalar processes in electroweak baryogenesis,''
  Phys.\ Rev.\  D {\bf 73}, 115009 (2006)
  [arXiv:hep-ph/0603058].

\bibitem{Fromme:2006wx}
  L.~Fromme and S.~J.~Huber,
  ``Top transport in electroweak baryogenesis,''
  JHEP {\bf 0703}, 049 (2007)
  [arXiv:hep-ph/0604159].


\bibitem{Kang:2004pp}
  J.~Kang, P.~Langacker, T.~j.~Li and T.~Liu,
  ``Electroweak baryogenesis in a supersymmetric U(1)' model,''
  Phys.\ Rev.\ Lett.\  {\bf 94}, 061801 (2005)
  [arXiv:hep-ph/0402086].

\bibitem{Carena:2004ha}
  M.~S.~Carena, A.~Megevand, M.~Quiros and C.~E.~M.~Wagner,
  ``Electroweak baryogenesis and new TeV fermions,''
  Nucl.\ Phys.\  B {\bf 716}, 319 (2005)
  [arXiv:hep-ph/0410352].

\bibitem{Huber:2006ma}
  S.~J.~Huber, T.~Konstandin, T.~Prokopec and M.~G.~Schmidt,
  ``Baryogenesis in the MSSM, nMSSM and NMSSM,''
  Nucl.\ Phys.\  A {\bf 785}, 206 (2007)
  [arXiv:hep-ph/0608017].

\bibitem{Fromme:2006cm}
  L.~Fromme, S.~J.~Huber and M.~Seniuch,
  ``Baryogenesis in the two-Higgs doublet model,''
  JHEP {\bf 0611}, 038 (2006)
  [arXiv:hep-ph/0605242].

\bibitem{Espinosa:2007qk}
  J.~R.~Espinosa and M.~Quiros,
  ``Novel effects in electroweak breaking from a hidden sector,''
  Phys.\ Rev.\  D {\bf 76}, 076004 (2007)
  [arXiv:hep-ph/0701145].

\bibitem{Bodeker:2004ws}
  D.~Bodeker, L.~Fromme, S.~J.~Huber and M.~Seniuch,
  ``The baryon asymmetry in the standard model with a low cut-off,''
  JHEP {\bf 0502}, 026 (2005)
  [arXiv:hep-ph/0412366].

\bibitem{Grojean:2004xa}
  C.~Grojean, G.~Servant and J.~D.~Wells,
  ``First-order electroweak phase transition in the standard model with a  low
  cutoff,''
  Phys.\ Rev.\  D {\bf 71}, 036001 (2005)
  [arXiv:hep-ph/0407019].

\bibitem{Delaunay:2007wb}
  C.~Delaunay, C.~Grojean and J.~D.~Wells,
  ``Dynamics of Non-renormalizable Electroweak Symmetry Breaking,''
  arXiv:0711.2511 [hep-ph].

%
\bibitem{Espinosa:1993yi}
  J.~R.~Espinosa, M.~Quiros and F.~Zwirner,
  ``On the electroweak phase transition in the minimal supersymmetric Standard
  Model,''
  Phys.\ Lett.\  B {\bf 307}, 106 (1993)
  [arXiv:hep-ph/9303317].

\bibitem{Espinosa:1996qw}
  J.~R.~Espinosa,
  ``Dominant Two-Loop Corrections to the MSSM Finite Temperature Effective
  Nucl.\ Phys.\  B {\bf 475}, 273 (1996)
  [arXiv:hep-ph/9604320].

%
\bibitem{de Carlos:1997ru}
  B.~de Carlos and J.~R.~Espinosa,
  ``The baryogenesis window in the MSSM,''
  Nucl.\ Phys.\  B {\bf 503}, 24 (1997)
  [arXiv:hep-ph/9703212].

\bibitem{Bodeker:1996pc}
  D.~Bodeker, P.~John, M.~Laine and M.~G.~Schmidt,
  ``The 2-loop MSSM finite temperature effective potential with stop
  condensation,''
  Nucl.\ Phys.\  B {\bf 497}, 387 (1997)
  [arXiv:hep-ph/9612364].

\bibitem{Cline:1998hy}
  J.~M.~Cline and G.~D.~Moore,
  ``Supersymmetric electroweak phase transition: Baryogenesis versus
  experimental constraints,''
  Phys.\ Rev.\ Lett.\  {\bf 81}, 3315 (1998)
  [arXiv:hep-ph/9806354].

\bibitem{Davidson:1999ii}
  S.~Davidson, T.~Falk and M.~Losada,
  ``Dark matter abundance and electroweak baryogenesis in the CMSSM,''
  Phys.\ Lett.\  B {\bf 463}, 214 (1999)
  [arXiv:hep-ph/9907365].

\bibitem{Laine:1998vn}
  M.~Laine and K.~Rummukainen,
  ``A strong electroweak phase transition up to $m_H\approx 105$ GeV,''
  Phys.\ Rev.\ Lett.\  {\bf 80}, 5259 (1998)
  [arXiv:hep-ph/9804255].

\bibitem{Losada:1996ju}
  M.~Losada,
  ``High temperature dimensional reduction of the MSSM and other  multi-scalar
  models,''
  Phys.\ Rev.\  D {\bf 56}, 2893 (1997)
  [arXiv:hep-ph/9605266].

\bibitem{Farrar:1996cp}
  G.~R.~Farrar and M.~Losada,
  ``SUSY and the electroweak phase transition,''
  Phys.\ Lett.\  B {\bf 406}, 60 (1997)
  [arXiv:hep-ph/9612346].

\bibitem{Csikor:2000sq}
  F.~Csikor, Z.~Fodor, P.~Hegedus, A.~Jakovac, S.~D.~Katz and A.~Piroth,
  ``Electroweak phase transition in the MSSM: 4-dimensional lattice
  simulations,''
  Phys.\ Rev.\ Lett.\  {\bf 85}, 932 (2000)
  [arXiv:hep-ph/0001087].


\bibitem{Laine:2000rm}
  M.~Laine and K.~Rummukainen,
  ``Two Higgs doublet dynamics at the electroweak phase transition: A
  non-perturbative study,''
  Nucl.\ Phys.\  B {\bf 597}, 23 (2001)
  [arXiv:hep-lat/0009025].


\bibitem{Cline:2006ts}
  J.~M.~Cline,
  ``Baryogenesis,''
  arXiv:hep-ph/0609145.

\bibitem{Quiros:2007zz}
  M.~Quiros,
  ``Electroweak baryogenesis,''
  J.\ Phys.\ A  {\bf 40}, 6573 (2007).

\bibitem{Buchmuller:2007fd}
  W.~Buchmuller,
  ``Baryogenesis -- 40 Years Later,''
  arXiv:0710.5857 [hep-ph].

\bibitem{Shaposhnikov:1986jp}
  M.~E.~Shaposhnikov,
  ``Possible Appearance of the Baryon Asymmetry of the Universe in an
  Electroweak Theory,''
  JETP Lett.\  {\bf 44}, 465 (1986)
  [Pisma Zh.\ Eksp.\ Teor.\ Fiz.\  {\bf 44}, 364 (1986)].

\bibitem{Quiros:1999jp}
  M.~Quiros,
  ``Finite temperature field theory and phase transitions,''
  arXiv:hep-ph/9901312.

\bibitem{Casas:2004uu}
  J.~A.~Casas, J.~R.~Espinosa and I.~Hidalgo,
  ``A relief to the supersymmetric fine tuning problem,''
  arXiv:hep-ph/0402017.

\bibitem{Cohen:1991iu}
  A.~G.~Cohen, D.~B.~Kaplan and A.~E.~Nelson,
  ``Spontaneous baryogenesis at the weak phase transition,''
  Phys.\ Lett.\  B {\bf 263}, 86 (1991).

\bibitem{Abel:1992za}
  S.~A.~Abel, W.~N.~Cottingham and I.~B.~Whittingham,
  ``Spontaneous baryogenesis in supersymmetric models,''
  Nucl.\ Phys.\  B {\bf 410}, 173 (1993)
  [arXiv:hep-ph/9212299].

\bibitem{Huber:1999sa}
  S.~J.~Huber, P.~John, M.~Laine and M.~G.~Schmidt,
  ``CP violating bubble wall profiles,''
  Phys.\ Lett.\  B {\bf 475}, 104 (2000)
  [arXiv:hep-ph/9912278].

\bibitem{Comelli:1993ne}
  D.~Comelli, M.~Pietroni and A.~Riotto,
  ``Spontaneous CP violation and baryogenesis in the minimal supersymmetric
  Standard Model,''
  Nucl.\ Phys.\  B {\bf 412}, 441 (1994)
  [arXiv:hep-ph/9304267].

\bibitem{Martin:1997ns}
  S.~P.~Martin,
  ``A supersymmetry primer,''
  arXiv:hep-ph/9709356.

\end{thebibliography}
\end{document}